\documentclass[prd,twocolumn,nofootinbib,showpacs,eqsecnum,floatfix,preprintnumbers,amsmath,amssymb]{revtex4}
\usepackage{graphicx}
\usepackage{dcolumn}
\usepackage{bm}

\usepackage{amsmath}
\usepackage{graphicx}
\usepackage{bbm}
\usepackage{color}
\usepackage{bm}
\usepackage{latexsym}
\usepackage{amssymb}
\usepackage{booktabs}
\usepackage[ansinew]{inputenc}
\usepackage{rotating}
\usepackage{euscript}
\usepackage{gensymb} 
\usepackage{subfigure}

\newcommand{\be}{\begin{equation}}

\begin{document}
\preprint{}

\title{Bosonic stringlike behavior and the ultraviolet filtering of QCD}

\author{Ahmed S. Bakry, Derek B. Leinweber, Anthony G. Williams}
\affiliation{Special Research Center for the Subatomic Structure of Matter, School of Chemistry \& Physics, University of Adelaide, South Australia 5005, Australia}
\email[]{abakry@physics.adelaide.edu.au}
\date{\today}


\begin{abstract}
  The gluonic action density is calculated in static mesons at finite temperature just below the deconfinement point. Our focus is to elucidate the role of vacuum ultraviolet  fluctuations which are filtered using an improved smearing algorithm. In the intermediate source separation distance, where the free string picture poorly describes the flux tube width profile, we find upon reducing the vacuum action towards the classical instanton vacuum, the characteristics of the flux tube converge and compare favorably with the predictions of the free bosonic string. This result establishes a connection between the free string action and vacuum gauge fields and reveals the important role of ultraviolet physics in understanding the lattice data at this temperature scale. As a by-product of these calculations, we find the broadening of the QCD flux tube to be independent of the ultraviolet filtering at large distances. Our results exhibit a linearly divergent pattern in agreement with the string picture predictions.   
\end{abstract}

\maketitle

\section{\label{sec:level1}Introduction}
  In the dual superconductor scenario of quark confinement, the quantum chromodynamical (QCD) vacuum squeezes the color fields into a confining string dual to the Abrikosov line by the dual Meissner effect. The string conjecture~\cite{luscherfr} follows as an intuitive realization of this squeezed color field with the major objective of deriving the leading and subleading properties of the flux tube in the infrared region of confining gauge theories. This effective description is expected to hold on distance scales larger than the intrinsic thickness of the flux tube $1/T_{c}$~\cite{Caselle:1995fh} in the rough phase of lattice gauge theories (LGT). The linearly rising potential part arises from the classical configuration of the string, and the quantum fluctuations of the string lead to the presence of a long distance $c/r$ term in the $q \overline{q}$ potential well known as the L\"uscher term. The existence of the subleading term has been verified in high precision measurements of Polyakov loop correlators in the SU(3) gauge group at zero temperature \cite{luscher}. The fluctuations of the string render an effective width for the flux-tube which grows logarithmically~\cite{Luscher:1980iy} as the color sources are pulled apart. The logarithmic divergence has been verified in many lattice simulations corresponding to a variety of confining gauge models~\cite{Bali,Pennanen:1997qm,Caselle:1995fh,Gliozzi:2010zt}. 

  At high temperature, higher-order gluonic modes are present. The corresponding free bosonic string predicts a new set of measurable thermal effects. These include a decrease in the effective string tension~\cite{Gao,deForcrand, Pisarski}, a change in the pattern of the tube's growth in width from a logarithmic divergence into a linear divergence~\cite{allais}, and a non-constant width profile~\cite{allais,PhysRevD.82.094503} along the $q \overline{q}$ line.

  Unlike the situation at zero temperature, the thermal behavior of the free string manifests only at source separation distance scales larger than what one expects normally in the zero temperature regime~\cite{Kac,allais,PhysRevD.82.094503,HariDass2008273}. The fact that the lattice data are poorly described by the free theory in the intermediate distance regime has been a subject of analytic and numerical studies which include higher-order terms of the effective string's action~\cite{Aharony:2009gg,Luscher:2004ib} into the corresponding partition function. The consequences of such an approach have been studied on the level of the $q\overline{q}$ potential~\cite{Caselle:818185,caselle2002} and, recently, has been extended to the flux-tube width profile~\cite{Gliozzi:2010zv,Gliozzi:2010zt,Caselle:2010zs}. Other studies investigate a possible finite intrinsic thickness of the QCD flux-tube~\cite{Vyas}.   

   Apart from the linearly rising potential, the interesting physics of the effective confining string is mainly due to its quantum fluctuations. As we will see, remarkable features arise when the ultraviolet (UV) part of the fluctuations of the stringlike flux tube is filtered out for intermediate quark separations at high temperatures. At this distance scale it is not yet clear if the deviations from the string picture are due to a non-Nambu-Goto action or the fact that a stringlike behavior has not yet set in. Thus, it is interesting to address this problem in a variant context by reporting an observation regarding the role played by the UV fluctuations of the vacuum in these discrepancies. We do this by tracking the response of the QCD vacuum, which is subject to UV filtering, to the presence of external static color sources. This work extends the region for which the free string picture is of utility.

  In the following we measure the gluonic action-density distribution by correlating an action density operator to Polyakov loop correlators. Measurements are taken on a set of SU(3) pure gauge configurations. The configurations are generated using the standard Wilson gauge action $S_{w}$ on two lattices of a spatial volume of $36^{3}$ and temporal extents of  $N_{t}=10$ and ${N}_{t}=8$, corresponding to temperatures $T/T_c \approx 0.8$ and $T/T_c \approx 0.9$ respectively. The simulations are performed for coupling value $\beta = 6.00$. At this value the lattice spacing is $a=0.1$ fm to reproduce the standard value of the string tension $\sqrt{\sigma}=440$ Mev~\cite{PhysRevD.47.661}. The Monte Carlo updates are implemented with a pseudo-heatbath algorithm~\cite{Cabibbo} using Fabricius-Haan and Kennedy-Pendelton (FHKP)~\cite{Fabricius,Kennedy} updating. Each update step consists of one heat bath and 5 over-relaxations. The measurements are taken on 500 bins separated with 2000 updating sweeps. Averaging inside each bin is performed by taking 5 measurements separated by 70 updating sweeps. This leads to a hierarchical integration, that is apart from updating the last time slice, similar to implementing a one-level L\"uscher Weisz (LW) algorithm~\cite{luscher}. 

  The measurements are taken after smoothing the gauge field by an over-improved stout-link smearing algorithm~\cite{Moran}. The value of the smearing parameters used are $\epsilon = -0.25$ and  $\rho_\mu = \rho = 0.06 $. Smoothing the gauge field reduces the action towards the action minimum or the classical instanton solution~\cite{deForcrand:2006my}. The UV characteristics of the gauge fields can be characterized in terms of Dirac eigenmodes. For example, the number of over-improved stout-link smearing sweeps used here has been calibrated to a given spectral cut-off $\lambda_{\rm{cutoff}}$ in the spectral representation of the Dirac operator~\cite{PhysRevD.77.074502}. The measurements are taken on sets of smeared gauge configurations with increasing levels of smearing. This way we are able to set the limit where the QCD vacuum response to the presence of an external static color sources asymptotically approaches the low energy free effective theory behavior.

\section{Quark--antiquark potential}
  At fixed temperature $T$, the Monte Carlo evaluation of the quark--antiquark potential at each $R$ is calculated through the Polyakov loop correlators    

\begin{align}
\label{Cor}
\mathcal{P}_{2Q} =& \int d[U] \,P(0)\,P^{\dagger}(R)\, \mathrm{exp}(-S_{w}),  \notag\\
                                       =& \quad\mathrm{exp}(-V(R,T)/T).
\end{align}
with the Polyakov loop given by

\begin{equation}
  P(\vec{r}_{i}) = \frac{1}{3}\mbox{Tr} \left[ \prod^{N_{t}}_{n_{t=1}}U_{\mu=4}(\vec{r}_{i},n_{t}) \right],
\end{equation}
 
  In the string picture, the Polyakov loop correlator assumes the functional form of the partition function of the two-dimensional bosonic string %

\begin{equation}
  \langle P(0) \,P^{\dagger}(R) \rangle= \int_{{\cal C}} [D\, X ] \,\mathrm{exp}(\,-S( X )).
\end{equation}  

\noindent The vector $X^{\mu}(\zeta_{1},\zeta_{2})$  maps the region ${\cal C}\subset R^{2}$ into $R^{4}$, with Dirichlet boundary condition $ X(\zeta_{1},\zeta_{2}=0)= X(\zeta_{1},\zeta_{2}=R)=0$, and periodic boundary condition along the time direction $ X (\zeta_{1}=0,\zeta_{2})= X (\zeta_{1}=L_T,\zeta_{2})$, $L_T=\frac{1}{T}$. $S$ is the string action in the physical gauge~\cite{luscher}

\begin{equation}
\label{path}
  S[ X ]=\sigma \, \frac{R}{T} + \frac{\sigma}{2}\int_{0}^{L_T} d\zeta_{1} \int_{0}^{R} d\zeta_{2} (\nabla X)^{2} + .....~ .
\end{equation}

\noindent The action decomposes into the classical configuration, the fluctuation part, and the string higher-order self interactions. A leading order approximation can be made by neglecting the self-interaction terms.
\noindent $\xi=(\xi_{1},i\xi_{2})$ is a complex parametrization of the world sheet, such that $\xi_{1}\in [-R/2,R/2], \xi_{2} \in [-L_T/2,L_T/2]$, and $\tau=\frac{L_T}{R}$ is the modular parameter of the cylinder. Solving the path integral of Eq.~\eqref{path} ~\cite{PhysRevD.27.2944}, and using Eq.~\eqref{Cor}, the quark anti-quark potential reads  

\begin{equation}
\label{sp}
 V(R,T)= 2\, T\, \log \eta \left(\frac{i}{2 T R}\right)+ \sigma R+ \mu(T).
\end{equation}  

\noindent $\eta$ is  the Dedekind eta function 

\begin{equation}
 \eta(\tau)=q^{\frac{1}{24}} \prod_{n=1}^{\infty}(1-q^{n});\quad  q=e^{\frac{-2\pi}{T\,R}},
\end{equation}
 and $\mu(T)$ is a renormalization parameter. 

   The numerical evaluation of the quark anti-quark potential, Eq.~\eqref{Cor}, using a four dimensional smearing scheme leads to a systematic ambiguity in regard to the transfer matrix interpretation which allows one to identify the expectation values of the Polyakov loop correlators with $\exp(-V(R,T)/T)$. We recourse, instead, to three dimensional smearing keeping the temporal-links unsmeared. The same smearing parameters as above are used. 

\begin{figure}[!hpb]
\begin{center}
\includegraphics[width=5.4cm] {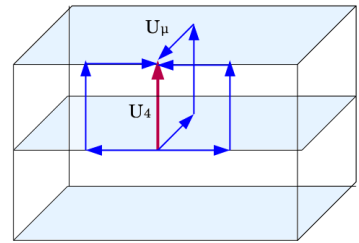}
\caption{ \label{newfig} The temporal link $U_4$ is updated based on the neighboring links. The shaded area represents the 3D spatial smeared lattice. The heat bath starts from smooth spatial links.}
\end{center}
\end{figure} 

\begin{figure*}[!hpt]
\begin{center}
\subfigure[~3D smearing.]{\includegraphics[width=8cm] {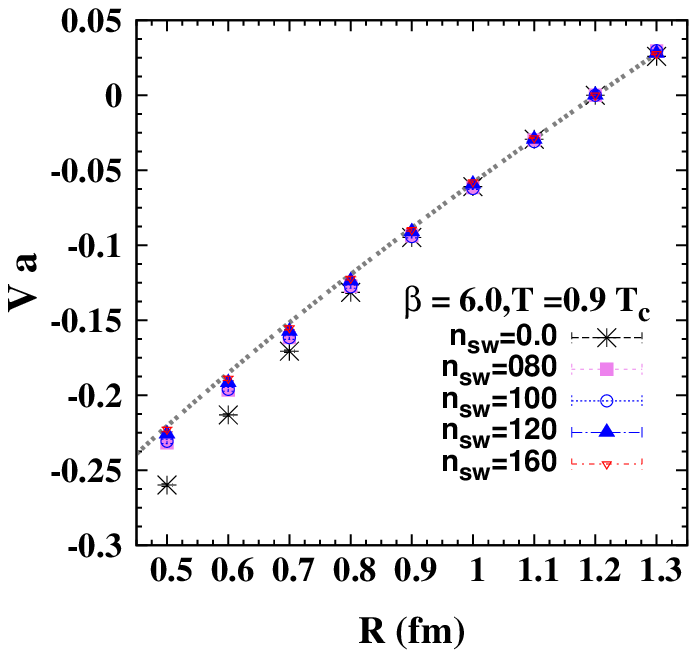}}
\subfigure[~4D smearing.]{\includegraphics[width=8cm] {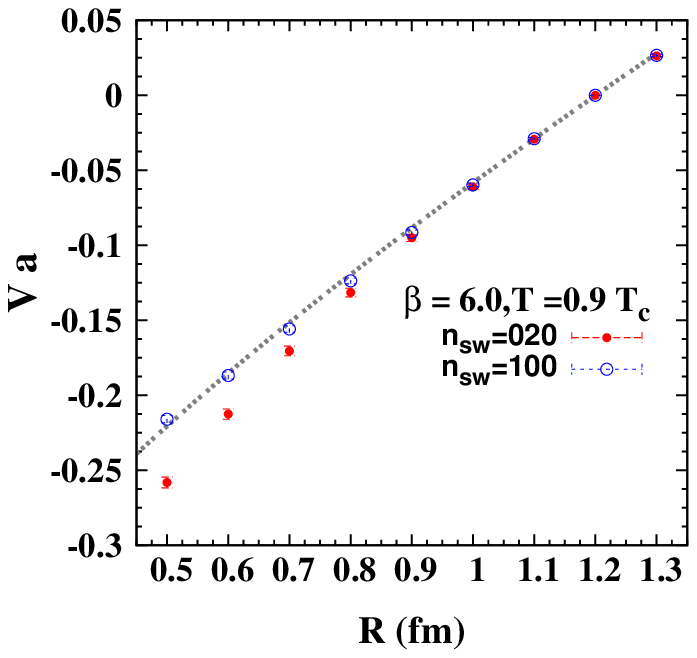}}
\end{center}
\caption{ \label{potfor} The quark--antiquark potential measured at each depicted smearing level for three and four-dimensional smearing, the lines correspond to the string picture predictions of Eq.~\eqref{sp}. The standard value of the string tension is used.}
\end{figure*}
\newpage
\noindent Our approach is as follows:
\begin{itemize}
  
\item  We start with 500 configurations. Each configuration is smeared only altering the spatial directions of the links, using spatially-oriented staples. This is depicted by shading of spatial links in Fig.~\ref{newfig}.  

\item For every 3D smeared configuration, an update sweep is applied. Each update sweep consists of a heat-bath step and four overrelaxation steps on the lattice. Four or more overrelaxation steps provide the same result within errors. Each step  proceeds as an update of every  single link based on its neighbors, effectively one at a time as one sweeps across the lattice in all four directions; spatial and temporal as indicated in Fig.~\ref{newfig}.  

\item An update sweep and subsequent measurement is repeated three more times on each configuration of the ensemble. This results in four measurements one over each of the four ensembles, where the ensembles differ in how many update sweeps they have had after spatial smearing, i.e, 1 to 4. 

  The measurements proceed as follows:

\item The temporal links in each of theses newly created configurations are integrated out using a source $Q$ sum of staples. The temporal link variables $U_{t}$ are replaced with the new link variable 

\begin{align}
\label{LI}
\bar{U_t}=\frac{\int dU U e^{-Tr(Q\,U^{\dagger}+U\,Q^{\dagger}) }}{\int dU e^{-Tr(Q\,U^{\dagger}+U\,Q^{\dagger}) }}.
\end{align}
     
\noindent using the numerical link-integration method of Ref.~\cite{1985PhLB15177D}. It should be noted, that unlike the link integration implementation of Ref.~\cite{Parisi} where pseudo-heat bath hits are performed only on the temporal links, the pseudo-heat bath updates we described above are performed on both the smeared space oriented links and the time oriented links. The temporal links, however, are integrated out by the numerical evaluation of the equivalent contour integral of Eq.~\eqref{LI} as detailed in Ref.~\cite{1985PhLB15177D}. 
 
\item Finally, the Polyakov loop correlators are calculated on each of the four one update sweep separated configurations. The result is averaged and binned as a single jackknife entry to avoid artificial error reduction. The 500 decorrelated bins are then averaged.       
\end{itemize}

  The Monte Carlo update step starts from a low action configuration in the spatial directions due to smearing. The above described update procedure brings in a newly updated time-link such that the effects of local action reduction that was only in the spatial torus takes place in the four-dimensional lattice. In this way, the UV filtering is implemented keeping the integration over the path integral Eq.~\eqref{Cor} systematic, thus, preserving the transfer matrix interpretation.

  In Fig.~\ref{potfor} the value of the potential measured on various levels of spatially smeared configurations, normalized to its value at $R=1.2$ fm, are plotted. Fig.~2(a) shows the numerical behavior of the data using the above described 3D smeared heat bath/overrelaxation driven updates. On the other hand, Fig.~2(b) shows the corresponding numerical behavior of the data measured on standard four-dimensional smeared configurations. The data corresponding to the unsmeared lattice and the string model predictions of Eq.~\eqref{sp} at $T/T_c \approx 0.9$ are also included. 

  The discrepancies between the unsmeared lattice data and the free string model occur in the intermediate distances $ R \leq  1$ fm. The numerical results for the quark--antiquark potential evaluated on the 3D smeared updated configurations show an interesting behavior with respect to the number of smearing sweeps (see Fig.~2(a)). The data at large distances show no response to the filtering of the UV fluctuations of the gauge field. The data at intermediate separation distances converge for a large number of smearing sweeps. Moreover, the data approach the free string model predictions.   

  It is interesting to compare these results to those obtained from the four-dimensional smearing illustrated in Fig.~2(b), where 20 and 100 sweeps of smearing are compared. The results for 100 sweeps of smearing coincide very well with the convergence toward the string model predictions observed in Fig.~2(a). Thus, the $4D$ smearing approach can be used as an efficient method for exploring the more demanding three-point functions required to determine the distribution of gluon flux, and this is used in the following. Similar results are observed at $T/T_c \approx 0.8$.    

\section{The Gluonic Profile}
  The transverse degrees of freedom of the string-like flux tube render an effective width for the tube. The mean-square width of the free bosonic string is defined as

\begin{align}
\label{operator}
   \omega^{2}(\xi;\tau) \equiv & \quad \langle \, X^{2}(\xi;\tau)\,\rangle \nonumber\\
                   = &\quad \dfrac{\int_{\mathcal{C}}\,[D\,X]\, X^2 \,\mathrm{exp}(-S[X])}{\int_{\mathcal{C}}[D\,X] \, \mathrm{exp}(-S[X])}.
\end{align}

  The above integral can be solved analytically \cite{allais,Gliozzi:2010zv}. The width of the tube in $D$ dimensions reads

\begin{equation}
\label{sol}
   \omega^{2}(\xi,\tau) = \frac{D-2}{2\pi\sigma}\log\left(\frac{R}{R_{0}(\xi)}\right)+\frac{D-2}{2\pi\sigma}\log\left| \,\dfrac{\theta_{2}(\pi\,\xi/R;\tau)} {\theta_{1}^{\prime}(0;\tau)} \right|,
\end{equation}

\noindent where $\theta$ are Jacobi elliptic functions, and $R_{0}(\xi)$ is  the UV cutoff which has been generalized to be dependent on distances from the sources. This solution gives the mean square width at all the planes transverse to the quark-antiquark line, and  hence, is describing the topological shape of the fluctuating flux tube and its dependence on the temperature as well as its evolution  with the increase of color source separation. In the above formula, we assume a dependence of the ultraviolet cutoff, $R_0$, on the position of the transverse planes, since as we will see, this quantity assumes different values near the quark sources as we fit the above formula to lattice data.   

  Using a modular transform $\tau \to -1/\tau$ ~\cite{allais, Gliozzi:2010zv}, Eq.\eqref{sol}, at $R >>\frac{1}{T}$, in four dimensions~\cite{Caselle:2010zs} becomes         

\begin{equation}
\omega^{2}(R/2;T)=\frac {1} {\pi\sigma} \log\left(\frac{2\,R_0}{T}\right) + \frac{1}{2 \sigma}T R -\frac{1} {\pi\sigma} e^{-2\pi R T} 
\end{equation}
which indicates linear growth of the tube's width at large distance.  

\begin{figure}[!hpt]
\begin{center}
\includegraphics[height=8cm,angle=-90]{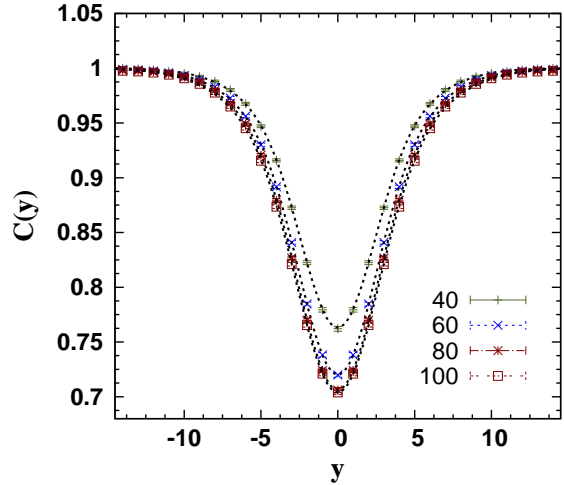}
\caption{\label{gauss}Plot of the density distribution $ \mathcal{C}(z=R/2,x_0,y)$ in the quark plane, $x_0$, at the center of the tube, $R/2$, for source separation  $R\,=0.9$ fm, $T/T_c \approx 0.9$.}
\end{center}
\end{figure} 

  The width of the action density of the free bosonic string can be compared to the width of the action density of the corresponding flux tube of the lattice gauge theory. After constructing the color-averaged infinitely-heavy static-mesonic state
\begin{align*}
\mathcal{P}_{2Q}(\vec{r}_{1},\vec{r}_{2}) =  P(\vec{r}_{1})P^{\dagger}(\vec{r}_{2}),
\end{align*}

\noindent subsequent measurement by an action density operator $\frac{1}{2}(E^{2}-B^{2})$ is taken at each point of the three-dimensional torus at each corresponding Euclidean time slice for every source configuration. The action density operator is constructed via a highly-improved $\mathcal{O}(a^{4})$ three-loop improved lattice field-strength tensor \cite{Bilson}. The measurements taken are averaged over the time slices.  A scalar field that characterizes the gluonic action-density distribution field can be then measured using the definition~\cite{Bissey}

\begin{equation}
\mathcal{C}(\vec{\rho};\vec{r}_{1},\vec{r}_{2} )= \frac{\langle\mathcal{P}_{2Q}(\vec{r}_{1},\vec{r}_{2}) \, S(\vec{\rho})\,\rangle } {\langle\, \mathcal{P}_{2Q}(\vec{r}_{1},\vec{r}_{2})\,\rangle\, \,\langle S(\vec{\rho})\, \rangle},
\label{Flux}
\end{equation}

\noindent where $< ...... >$ denotes averaging over configurations and lattice symmetries, and the vector $\vec{\rho}$ refers to the spatial position of the flux probe with respect to some origin. To further suppress the statistical fluctuations, the density distributions have been symmetrised around all the symmetry planes of the tube.

  Unless otherwise indicated, the measurements presented through out this section are taken on the 4D smeared configurations. The decrease in $\mathcal{C}(y)$ with the increase of the smearing sweeps is depicted in Fig.~\ref{gauss}.  The action-density asymptotically converges to a minimum for values around $n_{sw}=80$ to $100$ sweeps of smearing. Cluster decomposition of the operators leads to $C \rightarrow 1$ away from the quarks.

  A measurement of the width of the flux-tube's action density may be taken through fitting the density distribution $\mathcal{C}(\vec{\rho}(z,r,\theta))$, Eq.~\eqref{Flux}, to a Gaussian of the form 
\begin{equation}
\label{width1}
     \mathcal{C}(z;r,\theta)=1-a\,\exp[-r^2/\omega^2(z)]
\end{equation} 
  \noindent  with $r^2=x^2+y^2$ in each selected transverse plane $\vec{\rho}(z;r,\theta)$ to quark axis $z$ by making use of the cylindrical symmetry of the tube. 

  The mean square width of the flux tube is defined as the second moment of the flux density  with respect to the central line connecting the two quarks, $z$,

\begin{equation} 
\label{widthg}
\omega^{2}(z)=\quad \dfrac{\int \, dr\,r^{3}\,\exp[- r^{2} / \omega^{2}(z)  ]} {\int \,dr \,r \: \exp[ -r^{2}/\omega^{2}(z)] },
\end{equation}  

\noindent with  $z=0$ or $z=R$ denoting the position of the quark source. The width is normalized with respect to the ultraviolet cutoff $R_{0}(\xi)$ of Eq.~\eqref{sol} according to
\begin{equation}
  \omega_{n}^{2}(z)=\omega^{2}(z)+\frac{1}{\pi\,\sigma} \log(R_{0}(z)).
\end{equation}

\begin{figure}[!hpt]
\begin{center}
\includegraphics[width=8cm]{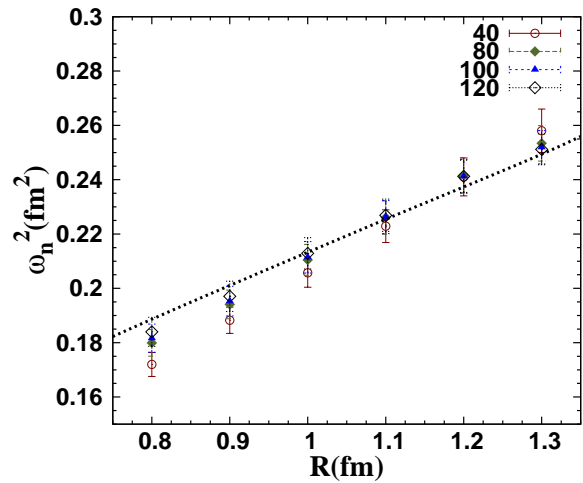} 
\caption{\label{mid} The mean square width of the flux tube $\omega_{n}^{2}(z=R/2)$ in the middle plane between the quarks. The lattice data, corresponding to the action density minimization, approach the string model predictions at short distances. At large distances the predicted linear divergence of the flux tube width is manifest in lattice data.}
\end{center}
\end{figure}

\begin{figure}[!hpt]
\begin{center}
\subfigure[Plane $z=2$.]{\includegraphics[width=8cm,angle=0]{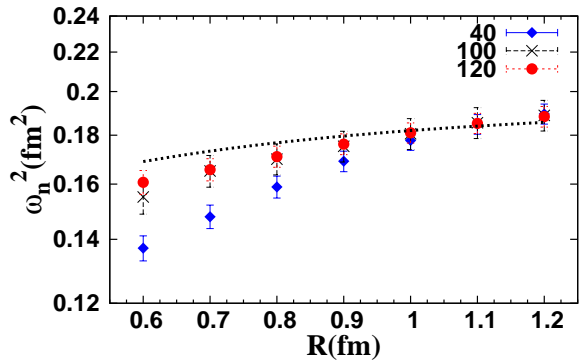}}\hfill
\subfigure[Plane $z=3$.]{\includegraphics[width=8cm,angle=0]{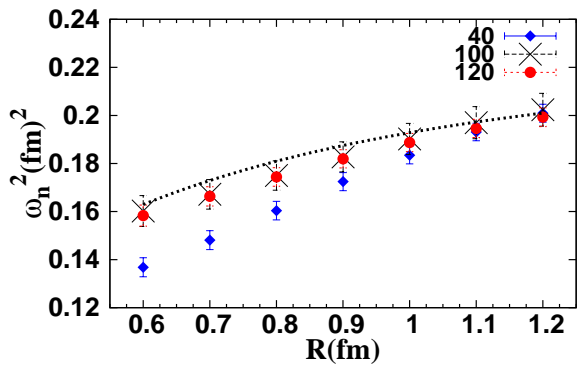}}\hfill
\subfigure[Plane $z=4$.]{\includegraphics[width=8cm,angle=0]{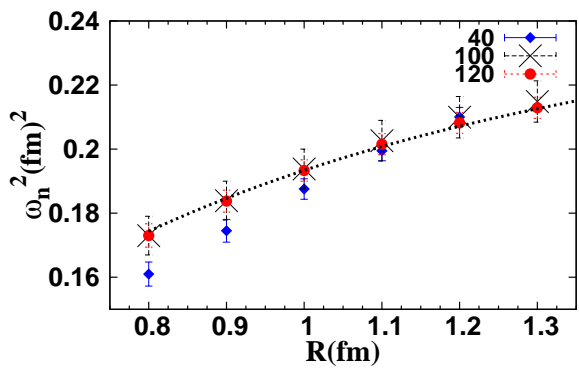}}
\caption{\label{planes}The normalized width  of the flux tube $ \omega_{n}^{2}(z) $ versus $q\overline{q}$ separations measured in the planes (a)~$z=2$,~(b)~$z=3$,~(c)~$z=4$, at ~$ T/T_c \approx 0.9$. The coordinates $z$ are lattice coordinates (lattice units) and are measured from the quark position $z=0$. The line denotes the one parameter string model, Eq.~\eqref{sol}, fit to lattice data for $R \geq 1$ fm. The numbers in the legend denote the number of smearing sweeps. }
\end{center}
\end{figure}

\begin{figure}[!hpt]
\begin{center}
\includegraphics[width=8cm]{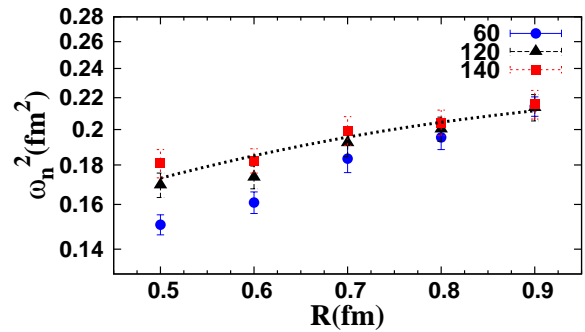} \hfill
\caption{\label{profilemh}Same as Fig.\ref{planes} for $\omega_{n}^{2}(z)$ measured at the plane $z=3$. The Polyakov lines are evaluated after integrating out the time links. The legend indicates the number of smearing sweeps applied before action measurements are taken.}
\end{center}
\end{figure}

  The measured values of the mean-square width in the middle plane $z_0=R/2$ of the tube versus the source separation are plotted in Fig.~\ref{mid}. Similar plots at three consecutive transverse planes $z=2$, $z=3$, and $z=4$ to the line joining the two color sources are illustrated in Figs.~\ref{planes} and ~\ref{profilemh}.
\noindent $R_{0}(\xi)$ has been measured for each smearing level and $\xi$ value by fitting Eq.~\eqref{sol} to data points having $R>1$ fm. Good $\chi^{2}$ is obtained.

  At large distances, in the middle plane of the flux tube, the tube shows a width broadening pattern for increasing $R$ that does not depend strongly on the corresponding smearing level. The data at large distances are increasing linearly in agreement with the string model predictions~\cite{Gliozzi:2010zv,Gliozzi:2010zt,Caselle:2010zs}. The UV effects, on the other hand, are manifest in the data points at shorter distances. The width of the flux-tube measured on the lowest smearing  level, where the short distance physics is best preserved, is poorly described by the free bosonic string model at short distances. As higher smearing levels are considered, the subsequent removal of the short distance physics from the gauge sector regulates the fast rate of growth of the flux tube width. However, this does not continue uncontrollably. The data ultimately converge near $100$ sweeps of smearing, in accord with the saturation in the action density of Fig.~\ref{gauss} and the potential in Fig.~2(a). Moreover, the UV-filtered results converge to the free string predictions.  

  Table~\ref{kai} summarizes the measured $\chi_{\rm{dof}}^{2}$ for fits of Eq.~\eqref{sol} for the fit range $R \geq 0.5$ fm. With the increase of the number of smearing sweeps the returned values of the $\chi_{\rm{dof}}^{2}$ improves and becomes stable near the regime of the action saturation of approximately 100 sweeps. 

\begin{table}[!hpt]
\caption{\label{kai} The returned $\chi^{2}_{\mathrm{dof}}$ by the fit of lattice data for measurement on the flux tube width in the middle plane between the quark anti-quark to the effective string model predictions of Eq.~\eqref{sol}. The lattice data correspond to smearing levels from $n_{sw}=40$ to $n_{sw}=120$}
\begin{center}
\begin{tabular}{cccccc}\hline
 No.Sweeps &$ 40 $& $60$ & $80$& $100$& $120$ \\\hline\hline
 $\chi_{\mathrm{dof}}^{2}$ &3.2 &1.6  &  1.20&0.98&0.96 \\\hline
\end{tabular}
\end{center}
\end{table}

\begin{figure}[!hpb]
\includegraphics[width=8cm]{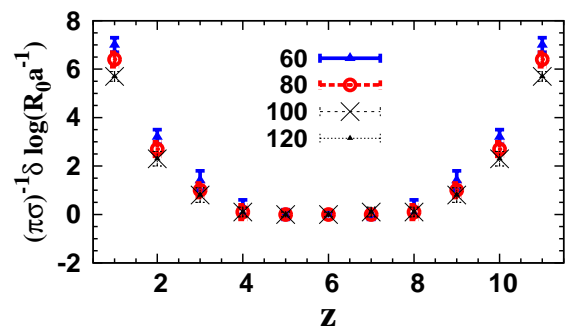}
\caption{\label{IT} The measured change in the ultraviolet cutoff $R_0$ along the flux tube. The smearing effect is small relative to the nontrivial dependence of $R_0$ on $z$.}
\end{figure}

\begin{figure}[!hpt]
\subfigure[$R=0.8\,\rm{fm}$] {\includegraphics[width=8cm]{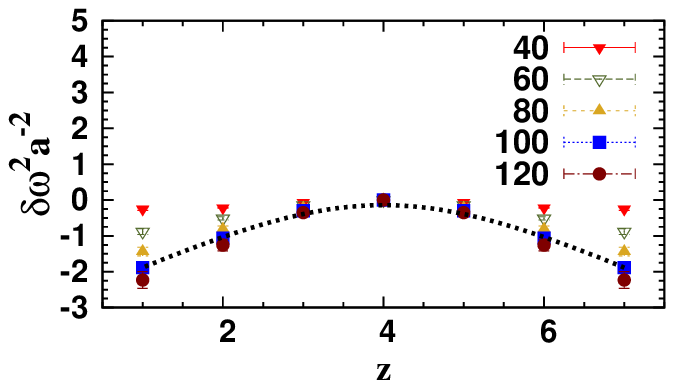}}\hfill
\subfigure[$R=0.9\,\rm{fm}$] {\includegraphics[width=8cm]{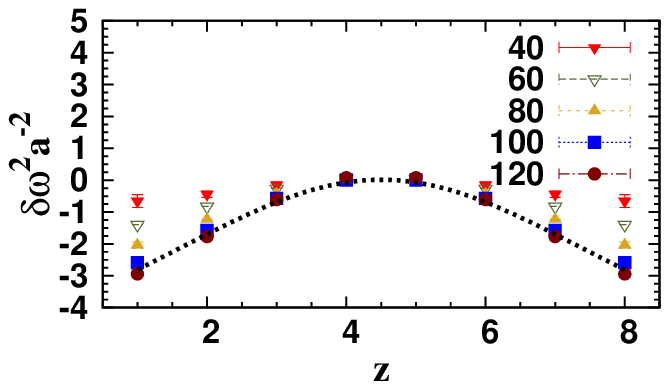}}\hfill
\subfigure[$R=1.2\,\rm{fm}$] {\includegraphics[width=8cm]{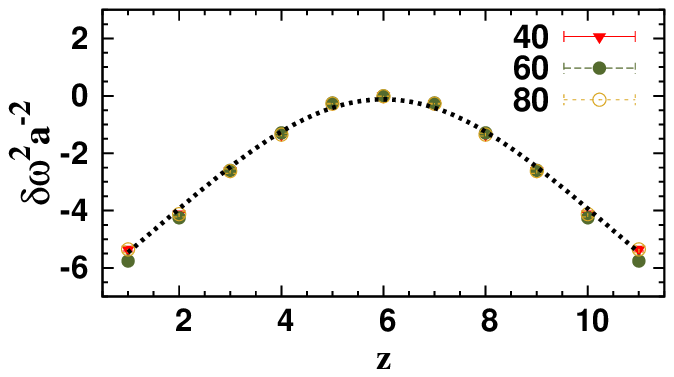}}
\caption{\label{profile} The change of the tube's width  $\delta \omega^{2}= \omega^{2}(z)-\omega^{2}(z_{0})$ measured from the central plane for the depicted $q\overline{q}$ separations. The smearing level of the lattice data is illustrated.  The line denotes the width difference $\delta\omega^{2}$ according to the string model Eq.~\eqref{sol}.~$\beta = ~6$, ~$ T/T_{c} \approx ~0.9$.}
\end{figure}

  To clarify this point further, we investigate the response of the QCD vacuum to the presence of infinitely heavy sources that are not constructed using smeared temporal links. Instead, the Polyakov lines are evaluated in the calculations of the flux strength, Eq.~\eqref{Flux}, using the numerical link integration~\cite{Parisi} procedure of Ref. ~\cite{1985PhLB15177D} for noise reduction.  The temporal links have not been smeared in the evaluation of Polyakov loops, rather the temporal links have been integrated out. This time, the Polyakov loops are taken from the unsmeared configurations and correlated with the smeared action density. For our analysis, performed on 500 configurations, we observe that the data corresponding to the width profile of the flux tube measured on high levels of vacuum UV filtering do display similar behavior to the results in Fig.~\ref{planes}. Fig.~\ref{profilemh} presents results for the plane $z=3$. Again the results systematically converge with a large number of smearing sweeps and approach the string model predictions. Note that the evaluation of the correlation function using this method involves a three point correlation function which becomes noisy at large distances.

  The shape aspects of the fluctuating free string are contained mainly in the second term of Eq.~\eqref{sol} and can be isolated by considering the difference in the mean square width at a given plane with respect to the central plane, $\delta \omega^{2}= \omega^{2}(z)-\omega^{2}(z_{0})$. The measured value of $R_0$, however, depends on the corresponding plane at which the lattice data is fit to Eq.~\eqref{sol}. The value of the fit parameter $R_0$ is fixed for each plane using lattice data at large separations $R=1.1$ fm and $R=1.2$ fm. 

  The changes in $R_0$ with respect to central plane is plotted separately in Fig.~\ref{IT}. The measurements performed at adjacent planes of the flux-tube reveal that this quantity varies along different planes orthogonal to the string. Indeed, we don't get a perfect match with the free-string profile unless we take such changes into account. The increase in the value of the UV cutoff $R_0$ is mostly obvious near the quarks and may indicate the importance of string self interactions near the boundary or the string interaction with the quark source itself. These interactions  are switched-off in the Nambu-Goto (NG) action and would appear when fitting the lattice data to the free string profile as a variation in the value of the UV cutoff along the tube. 

  This variation in $R_0$ has been included as part of the generalized string model solution of Eq.~\eqref{sol} illustrated by the curves in the following figures. A comparison of generalized string model predictions with the corresponding change in the flux tube's width on the lattice (obtained from fits of Eq.~\eqref{Flux}) is shown in Fig.~\ref{profile} for a source separation in the intermediate distance $R=0.8$ fm, $R=0.9$ fm and at large distance $R=1.2$ fm.

  Lattice data for each gauge smoothing level is also depicted in Fig.~\ref{profile}. The jackknife uncertainties associated with the change in the mean-squared width of the tube reveal correlated errors between the adjacent planes. Only subtle changes are observed in the tube's width along the transverse planes with respect to the central plane for the analysis performed on smeared gauge configurations of the lowest smearing level $n_{\rm{sw}}=40$ sweeps. The tube tends to exhibit larger curvatures as higher levels of gauge smoothing are considered. At large values of UV filtering, the tube profile converges and approaches the geometrical shape of the free-bosonic string. At large distances, on the other hand, the flux tube displays a curved width profile which compares well with the bosonic string profile and is not affected by smearing, as is evident in Fig.~\ref{profile} at $R=1.2$ fm. The string model curve has been calculated based on the changes in the second term of Eq.~\eqref{sol}.

\section{Conclusion}

   The presence of a pair of static external sources in the QCD vacuum induces a response of an effective free bosonic string for source separations in the intermediate separation region, provided the short distance vacuum fluctuations are filtered out. The flux tube, measured as a correlation between the mesonic operator and the vacuum action density,  is found to exhibit a broadening pattern  and a transverse structure similar to the free-bosonic string for measurements taken near the saturation in action density minimization under smearing even at intermediate distances. At large distances, the UV fluctuations do not affect the tube growth, which exhibits a linear divergent pattern consistent with the string model predictions. 

  In carrying out the $q\bar{q}$ potential calculations, we introduced a novel method for studying the effects of UV filtering of the QCD vacuum. This method avoids the ambiguities of performing unsystematic integrations due to smearing the temporal links, thus, preserving the transfer matrix interpretation. Instead three-dimensional spatial smearing is combined with single pseudo-heatbath driven updates. The numerical data of the $q\bar{q}$ potential obtained this way converge towards the string model predictions at large number of smearing sweeps. A comparison with four-dimensional smearing results reveals that any systematic effects associated with smearing the temporal links are subtle.

  The analysis performed at short distances provides an extension of the QCD vacua where the free string picture is of utility. The infrared region of the UV filtered vacuum can be described merely on the basis of a free string picture in the intermediate distances as well as large distances. This fact is relevant and complements recent investigations including higher-order self interactions to match lattice results. 

\section*{Acknowledgments}
 This research was undertaken on the NCI National Facility in Canberra, Australia, which is supported by the Australian Commonwealth Government.  We also thank eResearch SA for generous grants of supercomputing time which have enabled this project. This research is supported by the Australian Research Council.


\bibliography{Biblio}

\end{document}